\pdfoutput=1
\documentclass[11pt,twoside,a4paper,cmspaper,final,collab]{cms-tdr}
\def\svnVersion{66d7b1a}\def\svnDate{2024/06/15}\def\cmsCernNoTag{CERN-EP-2024-027}\def\cmsCernDate{\today}\def\cmsMessage{Published in Physical Review Letters as \href{http://dx.doi.org/10.1103/PhysRevLett.132.241802}{\doi{10.1103/PhysRevLett.132.241802}.}}
\begin{document}\cmsNoteHeader{TOP-22-003}

\cmsNoteHeader{TOP-22-003}
\newlength\cmsTabSkip\setlength{\cmsTabSkip}{1ex}
\newcommand{\emu}{\ensuremath{\Pepm\PGmmp}\xspace}
\newcommand{\mumu}{\ensuremath{\MM}\xspace}
\newcommand{\ee}{\ensuremath{\EE}\xspace}
\newcommand{\pp}{\ensuremath{\Pp\Pp}\xspace}
\newcommand{\tw}{\ensuremath{\PQt\PW}\xspace}
\newcommand{\ttzw}{\ensuremath{\ttbar\text{+\PW/\PZ}}\xspace}
\newcommand{\Wjets}{\ensuremath{\PW\text{+jets}}\xspace}
\newcommand{\xsectheo}{\ensuremath{832^{+20}_{-29}{\,\text{(scale)}} \pm 35\,({\mathrm{PDF}}+\alpS)\unit{pb}}\xspace}
\newcommand{\antikt}{anti-\kt}
\ifthenelse{\boolean{cms@external}}{\providecommand{\cmsLeft}{upper\xspace}}{\providecommand{\cmsLeft}{upper left\xspace}}
\ifthenelse{\boolean{cms@external}}{\providecommand{\cmsMiddle}{middle\xspace}}{\providecommand{\cmsMioddle}{upper right\xspace}}
\ifthenelse{\boolean{cms@external}}{\providecommand{\cmsRight}{lower\xspace}}{\providecommand{\cmsRight}{right\xspace}}

\title{Search for baryon number violation in top quark production and decay using   \texorpdfstring{proton-proton collisions at $\sqrt{s} = 13\TeV$}{sqrt(s)=13TeV}}

\author*[cern]{Reza Goldouzian and Mike Hildreth}
\date{\today}

\abstract{
A search is presented for baryon number violating interactions  in top quark production and decay.
The analysis uses data from proton-proton collisions at a center-of-mass energy of 13\TeV, collected with the CMS detector at the LHC with an integrated luminosity of 138\fbinv.
Candidate events are selected by requiring two oppositely-charged leptons (electrons or muons)  and exactly one jet identified as originating from a bottom quark.
Multivariate discriminants are used to separate the signal from the background. 
No significant deviation from the standard model prediction is observed. 
Upper limits are placed on the strength of baryon number violating couplings.  
For the first time the production of single top quarks via  baryon number violating interactions is studied.
This allows the search to set the most stringent constraints to date on the branching fraction of the top quark decay to a lepton, an up-type quark (\PQu or \PQc), and a down-type quark (\PQd, \PQs, or \PQb).  The results improve the previous bounds by three to six orders of magnitude based on the fermion flavor combination of the baryon number violating interactions.
}

\hypersetup{%
pdfauthor={CMS Collaboration},%
pdftitle={Search for baryon number violation in top quarkk production and decay using proton-proton collisions at sqrt(s) = 13 TeVs},%
pdfsubject={CMS},%
pdfkeywords={CMS, TOP, BNV, EFT}} % limit six total
\maketitle 

In the standard model (SM), the baryon number is a conserved quantum number. Its conservation, however, is not a direct consequence of fundamental symmetries within the SM, and it can be violated by nonperturbative effects \cite{tHooft:1976rip}.
The size of such violations is too small to explain the observed matter-antimatter asymmetry in the universe \cite{Sakharov:1967dj}.
Certain scenarios of physics beyond the SM, such as grand unified theories \cite{Georgi:1974sy} and supersymmetry \cite{Weinberg:1981wj}, naturally include baryon number violation (BNV) and could provide a mechanism to explain this observation.  Various low-energy direct searches for signatures of BNV have been conducted over the past decades, with constraints set on the BNV energy scale via processes such as nucleon \cite{Super-Kamiokande:2020wjk}, \PGt lepton \cite{LHCb:2013fsr}, \PQc \cite{BESIII:2021krj}, and \PQb quark \cite{BaBar:2011yks} decays. 
There are also stringent indirect constraints from proton stability involving heavy quarks \cite{Hou:2005iu}  for specific theoretical assumptions \cite{McCracken:2015coa}. 
Experiments at the CERN LHC provide the highest sensitivity for potential high-energy BNV processes involving the top quark.  Previously, the CMS Collaboration has performed a search for BNV decays of the top quark in single lepton (electron or muon) channels in proton-proton (\pp) collisions at  $\sqrt{s}$ =  8 \TeV \cite{CMS:2013zol}.  
This Letter presents the first search for top quark BNV interactions via  single top quark production in association with a lepton  in \pp collisions at 13\TeV  in dilepton final states.  
The data used in the analysis correspond to an integrated luminosity of 138\fbinv, collected by the CMS experiment at the LHC during 2016--2018. 

\begin{figure}[h!]
      \includegraphics[width=1\columnwidth]{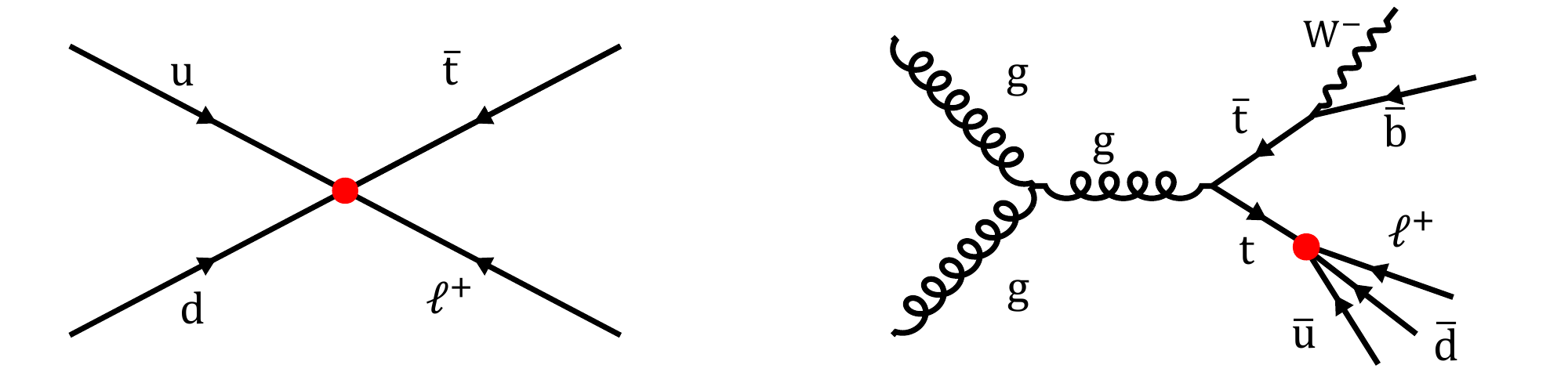}
    \caption{Representative Feynman diagrams for single top quark production (left) and  top quark decays (right) via BNV interactions. The red circles mark the BNV vertices.
    \label{Feyn}}
\end{figure}

Assuming the mass scale of new physics responsible for BNV processes is larger than the energy scale directly accessible at the LHC, BNV interactions of top quarks can be described through an effective Lagrangian, $\mathcal{L}_\text{eff}$. 
Assuming the SM field content and gauge symmetries, the terms in the BNV effective Lagrangian must involve three quark fields and one lepton field~\cite{Weinberg:1979sa}.
Including up to dimension-six operators, the most general effective Lagrangian that describes the BNV interactions of the top quark and a charged lepton takes the form \cite{Dong:2011rh}:

\begin{equation}
		 \mathcal{L}_\text{eff} =  \frac{C_s}{\Lambda^2} \epsilon^{\alpha\beta\gamma}
		[ \overline{\PQt^c_\alpha}  \PQd_\gamma	] 
		[ \overline{\PQu^c_\beta}  \Pell ] 
		+ \frac{C_t}{\Lambda^2}  \epsilon^{\alpha\beta\gamma}
		[\overline{\PQt^c_\alpha}  \Pell ]
		[\overline{\PQu^c_\beta}  \PQd_\gamma ] + \rm{h.c.},
	\label{eq1}
\end{equation}
where  \PQd,  \PQu, and \Pell  are down-type quark, up-type quark, and charged-lepton fields, respectively, where the superscript ``c'' denotes charged conjugated fields.
Colors are labeled by greek indices, $\Lambda$ is the generic scale of new physics,  and $C_s$ and $C_t$ are fermion-flavor-dependent effective couplings.
The $s$ and $t$ labels in Eq.~(\ref{eq1})  denote that the new physics scale may be linked to the mass of a heavy mediator exchanged in the $s$ or $t$ channels, respectively~\cite{Dong:2011rh}.  
No specific chirality is assumed for the BNV interactions.
The terms in  $\mathcal{L}_\text{eff}$ violate  baryon and lepton numbers simultaneously.
These effective interactions open new top quark decay and production channels at the LHC.
Figure~\ref{Feyn} displays representative Feynman diagrams for single top quark production (``ST mode") and  top quark decay (``TT mode")  via BNV interactions in top quark-antiquark pair production (\ttbar).

This analysis uses events in  dileptonic final states (\ee, \emu, and \mumu) where one lepton is produced via the BNV interaction and a second lepton comes from the decay of the \PW boson produced in the dominant $\PQt \to \PQb \PW$ decay.  The strength of the twelve flavor combinations of top quark four-fermion BNV interactions are probed in these final states. These take the form $\PQt \Pell \cPq_\PQu \cPq_\PQd$, where \Pell can be an electron or muon, $\cPq_\PQu$ can be an up or charm quark, and $\cPq_\PQd$ can be any down-type quark.
The BNV interactions with a tau lepton can contribute to the dileptonic final states considered in this analysis via its leptonic decays. However, only a small fraction of these events appears in the signal selection because of the low branching fraction of the leptonic final states  and the lower energies  of the decay electrons or muons. Therefore, $\PQt \Pgt \cPq_\PQu \cPq_\PQd$ interactions are not probed in this analysis.   

The central feature of the CMS apparatus is a superconducting solenoid of 6\unit{m} internal diameter, providing a magnetic field of 3.8\unit{T}.
Within the solenoid volume are a silicon pixel and strip tracker, a lead tungstate crystal electromagnetic calorimeter, and a brass and scintillator hadron calorimeter, each composed of a barrel and two endcap sections.
Forward calorimeters extend the pseudorapidity ($\eta$) coverage provided by the barrel and endcap detectors. Muons are measured in gas-ionization detectors embedded in the steel flux-return yoke outside the solenoid.
A more detailed description of the CMS detector, together with a definition of the coordinate system used and the relevant kinematic variables, can be found in Ref.~\cite{Chatrchyan:2008zzk}.

The data are modeled by Monte Carlo (MC) simulations of the signal and background processes.
Simulated events are produced with event generator programs using  the next-to-next-to-leading order (NNLO) parton distribution function (PDF) sets from NNPDF3.1 \cite{Ball:2017nwa}.
Parton showering and hadronization are done with \PYTHIA v8.240~\cite{Sjostrand:2014zea} using the underlying-event tune  CP5~\cite{Sirunyan:2019dfx}.
Generated events undergo a full simulation of the detector response using \GEANTfour~\cite{Agostinelli:2002hh}.
The presence of simultaneous \pp collisions in the same or nearby bunch crossings, referred to as pileup (PU), is modeled by superimposing inelastic \pp interactions, simulated using \PYTHIA, on all MC events.
Simulated events are then reweighted  to reproduce the PU distribution observed in data. 

Contributions to the background include SM \ttbar production, single top quark production in association with a \PW boson (\tw),  \PW or \PZ bosons produced in association with \ttbar (\ttzw),  Drell--Yan in association with jets (DY+jets) processes, \Wjets production, and  diboson processes (including $\PW\PW$, $\PW\PZ$, and $\PZ\PZ$). The contribution from quantum chromodynamics (QCD) multijet production is
found to be negligible.
The \POWHEG v2.0 next-to-leading order (NLO) MC generator  is used to simulate the SM \ttbar, \tw, and diboson events ~\cite{bib:powheg0,bib:powheg2,Frixione:2007vw}. 
The \MGvATNLO v.2.6.5 generator is used to simulate the \ttzw, DY+jets,  \Wjets, and   diboson events~\cite{Alwall:2014hca}.  
The \ttbar sample is normalized to the cross section calculated
at NNLO in QCD including resummation of next-to-next-to-leading logarithmic (NNLL) soft-gluon terms with the \textsc{Top++}2.0 program~\cite{Czakon:2011xx}, \xsectheo, where \alpS is the strong coupling.
To improve the modeling of the transverse momentum (\pt) spectrum of the top quark in \POWHEG, simulated SM \ttbar events are weighted as a function of the \pt of the top quark to match the expectations at NNLO QCD accuracy, including electroweak corrections~ \cite{Czakon:2017wor}.
Other simulated samples are normalized to their cross section predictions at NNLO (for DY+jets and \Wjets ~\cite{Li:2012wna}), NLO+NNLL (for \tw production~\cite{bib:twchan}), or NLO (for the diboson and \ttzw processes~\cite{bib:mcfm:diboson,bib:Maltoni:2015ena}).

The simulated signal samples are generated using the \MGvATNLO v.2.6.5 generator with the TopBNV model \cite{Dong:2011rh} at leading order in QCD.
The top quark BNV signal sample has two independent components: 
(i) ST mode,  and
(ii) TT mode,  as shown in Fig.~\ref{Feyn}.
Independent samples are generated for the various possible fermion flavor combinations. The top quark mass and width are set to 172.5 and 1.33 GeV, respectively. 
There is no interference considered between the BNV signal and SM processes and the BNV couplings  affect only the signal yield. Since the BNV couplings are probed individually, signal samples are generated separately for nonzero $C_t$ and $C_s$ couplings, assuming $\Lambda =1\TeV$. 
The BNV signal cross sections and the branching fractions for the BNV top quark decays depend quadratically on $C/\Lambda^2$ \cite{Dong:2011rh}.
Theoretical cross sections for single top quark production and top quark decays via the BNV interactions are shown in Table~\ref{tab:CrossSec}. The dominant signal process is the ST mode because of its larger cross section.
Final-state particles in the ST mode also have a harder \pt spectrum compared to SM processes and the TT mode~\cite{Dong:2011rh}.
Therefore,  the analysis is optimized with respect to the ST mode signatures and  the TT mode contribution is added for completeness.

\begin{table}[!ht]
\topcaption{Theoretical inclusive cross sections, in  units of pico barn (pb), for single top quark
production (ST) and top quark-antiquark pair production with the decay (TT) via BNV interactions, assuming a top quark mass of 172.5\GeV, the top quark decay width 1.33\GeV, $\Lambda=1\TeV$, and $C_t =1$ or $ C_s=1$. 
The uncertainties arising from the choice of the renormalization and factorization scales and PDFs are given as ($\sigma{\pm\text{Scale}}{\pm\text{PDF}}$). Here, the sum of the two cross sections is given where \Pell =  \Pe  or \Pgm.
\label{tab:CrossSec}}
\centering
\begin{scotch}{lll}
Process  & $\sigma(C_t=1)$ [pb] & $\sigma(C_s=1)$ [pb] \\
\hline
ST ($\PQt\Pell \PQu\PQd$) & $31.5 \pm 2.1 \pm 1.0$   & $10.7 \pm 0.7  \pm 0.4$    \\
[\cmsTabSkip]
ST ($\PQt\Pell \PQu\PQs$)  & $8.1 \pm 0.3 \pm 0.5$   & $2.8 \pm 0.1 \pm 0.2$    \\
[\cmsTabSkip]
ST  ($\PQt\Pell \PQu\PQb$)  & $3.31 \pm 0.13 \pm 0.06$    & $1.14 \pm 0.05 \pm 0.02$     \\
[\cmsTabSkip]
ST  ($\PQt\Pell \PQc\PQd$) &$2.77 \pm 0.22 \pm  0.01$   & $0.96 \pm 0.01  \pm 0.07$    \\
[\cmsTabSkip]
ST  ($\PQt\Pell \PQc\PQs$)  &$0.79 \pm  0.02 \pm  0.11$    & $0.27 \pm  0.01 \pm  0.04$     \\
[\cmsTabSkip]
ST  ($\PQt\Pell \PQc\PQb$) & $0.28 \pm  0.03 \pm  0.04$   & $0.10 \pm  0.01 \pm  0.01$    \\
[\cmsTabSkip]
TT  & $0.007 \pm 0.002 \pm 0.001$ & $0.007 \pm 0.002 \pm 0.001$   \\ 
\end{scotch}
\end{table}

Signal events in the ST mode contain two opposite-sign leptons,  one jet originating from a bottom quark, and missing energy due to the undetected neutrino from the top quark decay. 
Events were selected online during  data taking  by a combination of single-electron and di-electron triggers  for the \ee events as in Refs. \cite{CMS:2018fks,CMS:2016ngn}. 
Single-muon triggers are used for the \emu and  \mumu events , as described in Ref. \cite{CMS:2021ctt}.  
The particle-flow (PF) algorithm aims at reconstructing individual particles (photons, charged and neutral hadrons, muons, and electrons) by combining information from the various components of the CMS detector \cite{Sirunyan:2017ulk}.
The primary vertex is taken to be the vertex corresponding to the hardest scattering in the event, evaluated using tracking information alone,  as described in Section 9.4.1 of Ref.~\cite{CMS-TDR-15-02}.
Electron and muon candidates \cite{Khachatryan:2015hwa,Sirunyan:2018fpa} are 
required to lie within $|\eta|<2.4$ to keep them within the silicon tracker coverage.
Electron candidates in the transition region between barrel and endcap calorimeters ($1.44 < \abs{\eta} < 1.57$) are removed.
The same high-\pt  lepton identification and isolation criteria described in Ref. \cite{CMS:2021ctt} are used to reject nonprompt leptons. 
Events are required to have exactly two opposite-sign leptons.  To operate well above the trigger threshold,  the selected electron (muon) should  have $\pt>35 (53)\GeV$.  Selected events are divided based on their lepton flavors into three mutually exclusive categories:  \ee, \emu, and \mumu.
To suppress backgrounds, especially from the DY+jets processes, we reject events in which the two leptons have an invariant mass below 106\GeV. 

Jets are reconstructed from the PF candidates using the \antikt clustering algorithm with a distance parameter of 0.4 \cite{Cacciari:2008gp,Cacciari:2011ma}. We select jets with $ \abs{\eta} <2.4$ and $\pt>30\GeV$.
~Jets originating from \PQb quarks (\PQb jets) are identified (\PQb-tagged) using the  \textsc{DeepJet} algorithm \cite{Sirunyan:2017ezt} with an average \PQb tagging efficiency of 68\% and a light quark and gluon jet misidentification rate of 1.1\%. Events are required to have exactly one \PQb-tagged jet.
The missing transverse momentum (\ptvecmiss)  is defined as the negative vector \ptvec sum
of all PF particles, and its magnitude is denoted as \ptmiss~\cite{CMS:2019ctu}.  Events with \ptmiss $< 60$ \GeV are rejected to further suppress the DY+jets events.

The selected events in the signal region have exactly one opposite-sign lepton pair with invariant mass greater than $106\GeV$, $\ptmiss > 60\GeV$, and exactly one \PQb-tagged jet irrespective of the number of untagged jets.
The dominant background is the \ttbar process (${\sim}89\%$), followed by the \tw (${\sim}9\%$) and DY+jets (${\sim}1\%$)  processes. 
Signal selection efficiencies for the ST (TT) mode are about 2.8 (1.1)\% with respect to an inclusive MC sample  for the $\PQt\Pell \PQu\PQd$ flavor combination, assuming $C_s=0$ and nonzero $C_t$.  
The fractional uncertainties on the signal efficiencies due to the effects described below range from 4-9\%.

In the ST mode, the lepton and top quark are produced directly from the annihilation of the incoming quarks, and are Lorentz-boosted and approximately back-to-back. The subleading lepton in the ST mode is primarily from the top quark decay chain. To use these specific features of the signal events, the four-momentum vectors of the
top quarks are reconstructed from the decay products: the subleading lepton, the neutrino, and the \PQb jet candidate.
The neutrino \pt can be inferred from the \ptvecmiss.
The longitudinal momentum of the neutrino is inferred assuming energy-momentum conservation at the \PW boson decay vertex and constraining
the \PW boson mass to $80.4$ \GeV as  discussed in Ref.~\cite{CMS:2016lel}. 

A boosted decision tree (BDT)~\cite{Hocker:2007ht,QUINLAN1987221} is employed to distinguish signal from the sum of the background processes.
Independent signal and background samples are used for training the BDT.
A merged sample from \ttbar, \tw, and DY+jets background events, weighted by their cross sections, is used as the background training and testing sample.
For the signal, events from various BNV flavor combinations of the ST mode  are merged with equal weights and are used in the BDT training and testing.
Ten variables are inputs to the BDT: the transverse momenta of the leading lepton ($\Pell_1$), subleading lepton ($\Pell_2$), and the top quark candidate (t); the distances between the leading and subleading leptons [$\Delta R(\Pell_1, \Pell_2) = \sqrt{\smash[b]{( \eta^{\Pell_1} - \eta^{\Pell_2})^2 +  (\phi^{\Pell_1} - \phi^{\Pell_2})^2}}$] and $\Delta R(\Pell_1,t)$; the azimuthal angles between the leading and subleading leptons [$\Delta \phi(\Pell_1, \Pell_2) =|\phi^{\Pell_1} - \phi^{\Pell_2}|$] and $\Delta \phi(\Pell_1,\PQt)$; the invariant mass and \pt of the dilepton system; and $\abs{\pt^{\PQt}-\pt^{\Pell_1}}/\abs{\pt^{\PQt}+\pt^{\Pell_1}}$.

The templates describing the BDT distributions for the signal and background events are taken from simulation. 
The normalization of the DY+jets background, which is important in
the \ee and \mumu channels, is determined by applying a scale factor to the simulation derived from data in a control region where the reconstructed dilepton mass is close to the \PZ boson mass~\cite{CMS:2015yky}.

The list of uncertainties considered and the techniques used to estimate their values are very similar to those in Ref.~\cite{CMS:2022ztx}. 
We consider uncertainties in the integrated luminosity~\cite{CMS-LUM-17-003,LUM-17-004,LUM-18-002}, pileup effects \cite{Sirunyan:2018nqx}, trigger, lepton identification ~\cite{Khachatryan:2015hwa,Chatrchyan:2012xi}, and \PQb tagging \cite{Sirunyan:2017ezt,CMS-DP-2023-005} efficiencies, in the calculation of \ptmiss, and those related to the jet energy scale and resolution \cite{Khachatryan:2016kdb}.
Uncertainties arising from choices in signal and \ttbar modeling include PDFs,  renormalization and factorization scales, and initial- and final-state QCD radiation.    
The uncertainty arising from the modeling of the top quark \pt  spectrum is evaluated by the renormalization and factorization scales  at NNLO QCD accuracy,  including NLO  electroweak corrections~ \cite{Czakon:2017wor}.
Modeling uncertainties from the matching of the matrix element level calculation  to the parton shower simulation, the modeling of the underlying event defined in \PYTHIA tunes, and the models of color reconnection are considered for the SM \ttbar process,  as  described in Ref. ~\cite{Sirunyan:2018ucr}.
The modeling uncertainties apply only to the signal acceptances.
Normalization uncertainties of 5, 10, and 30\% are considered for the \ttbar, \tw,  and other processes based on experimental measurements \cite{Sirunyan:2018ucr,Sirunyan:2018lcp}, respectively.
An additional 20\% normalization uncertainty is added for DY+jets processes to account for PU mismodeling  in large \ptmiss events.

\begin{figure}[htb!]
  \begin{center}
      \includegraphics[width=0.48\textwidth]{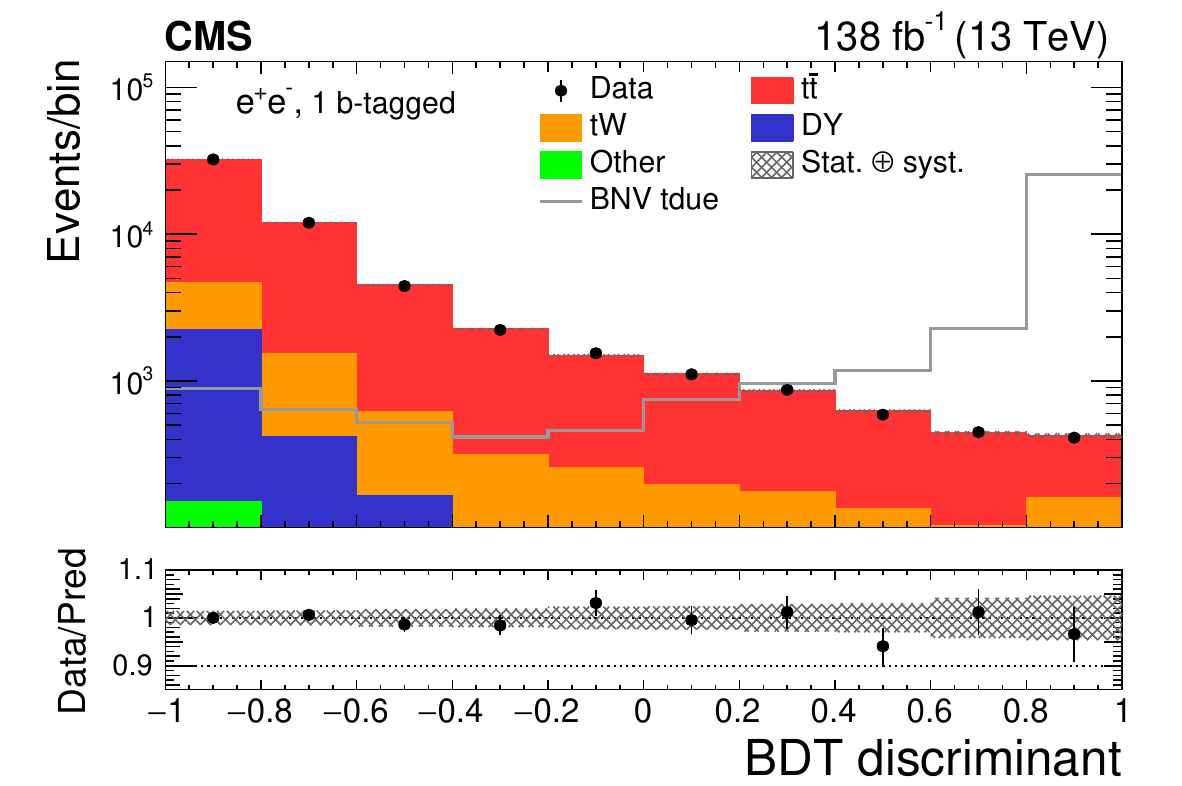}
      \includegraphics[width=0.48\textwidth]{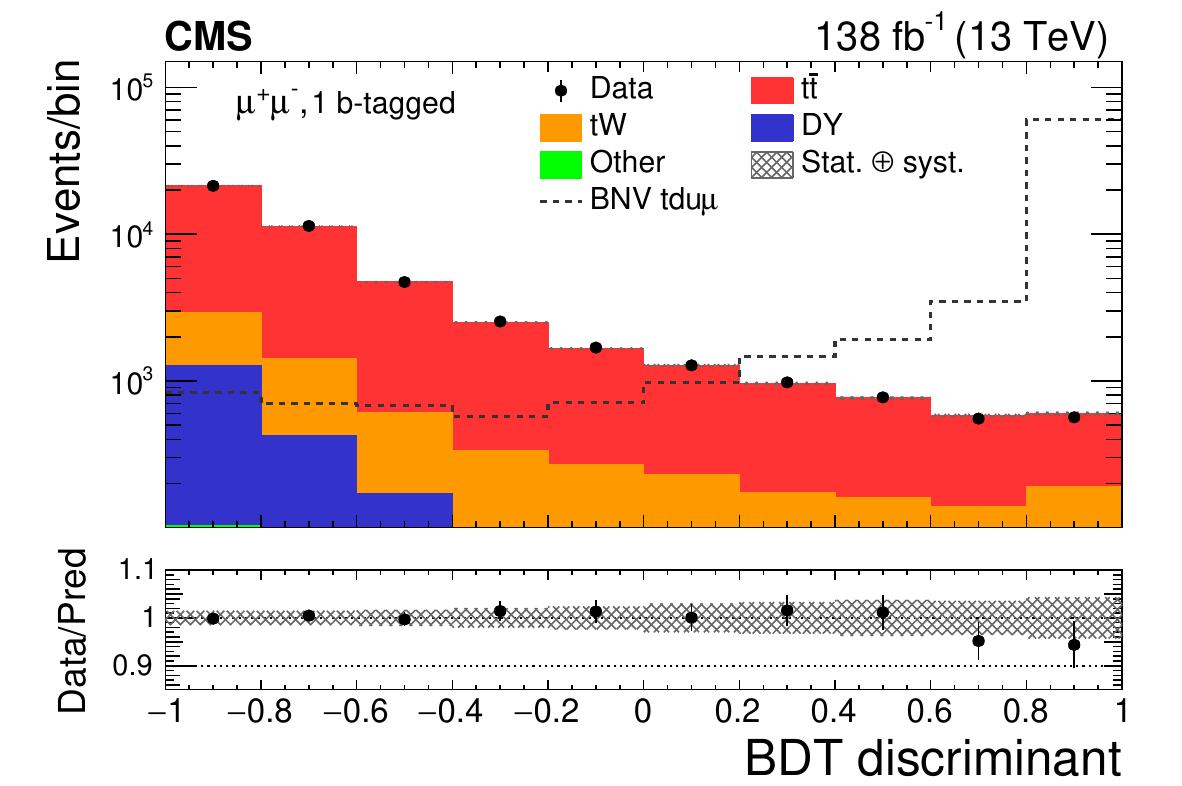}
      \includegraphics[width=0.48\textwidth]{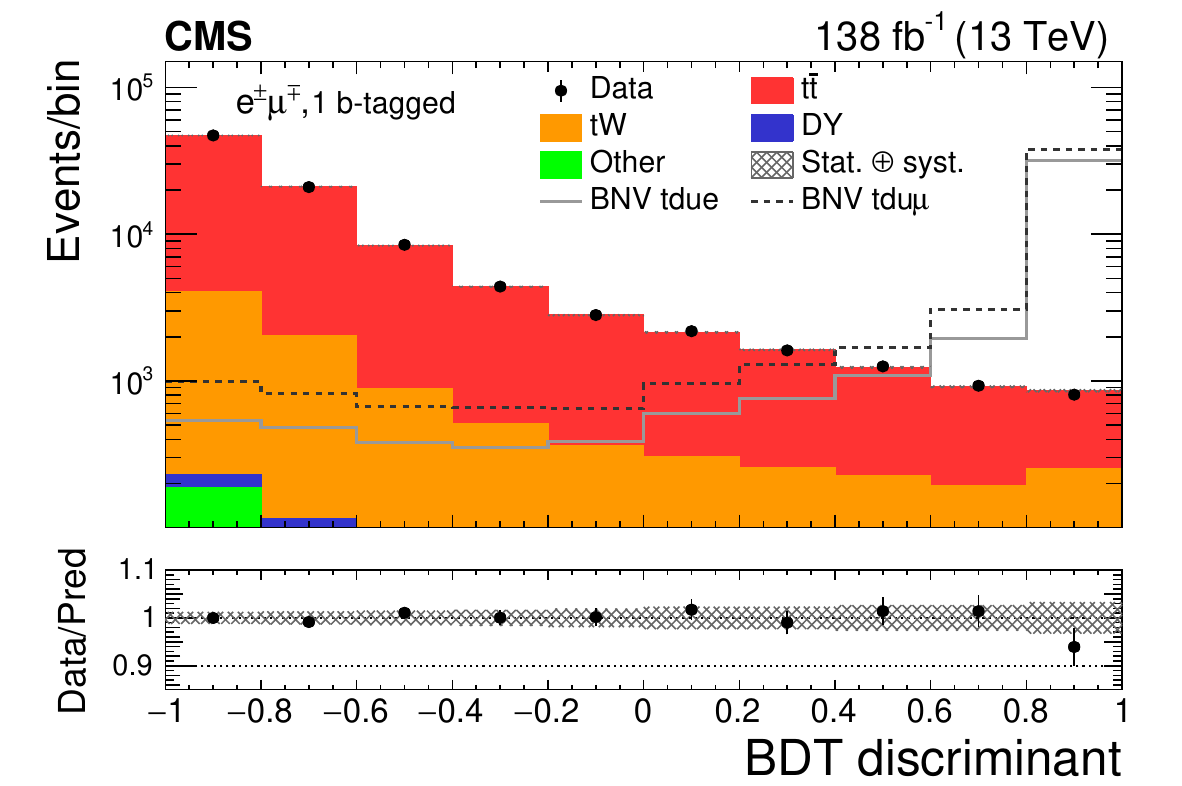}
  \caption{The BDT output distributions for data (points) and backgrounds (histograms)  for the \ee (\cmsLeft),  \mumu (\cmsRight), and \emu (lower) channels, including the ratio of data to the
predicted total background yield.
The hatched bands indicate the total uncertainty (statistical and systematic added in quadrature) for the SM background predictions.
The predicted yields of the backgrounds and the uncertainty bands are shown after the simultaneous fits for the signal-plus-background hypothesis. 
Examples of the predicted signal contribution for the BNV interactions via $\PQt\Pe\PQu\PQd$ (solid gray line) and $\PQt\PGm\PQu\PQd$ (dashed black line) vertices are shown.
}
  \label{fig:BDT}
  \end{center}
\end{figure}

\begin{figure}[htbp]
  \begin{center}
      \includegraphics[width=0.48\textwidth]{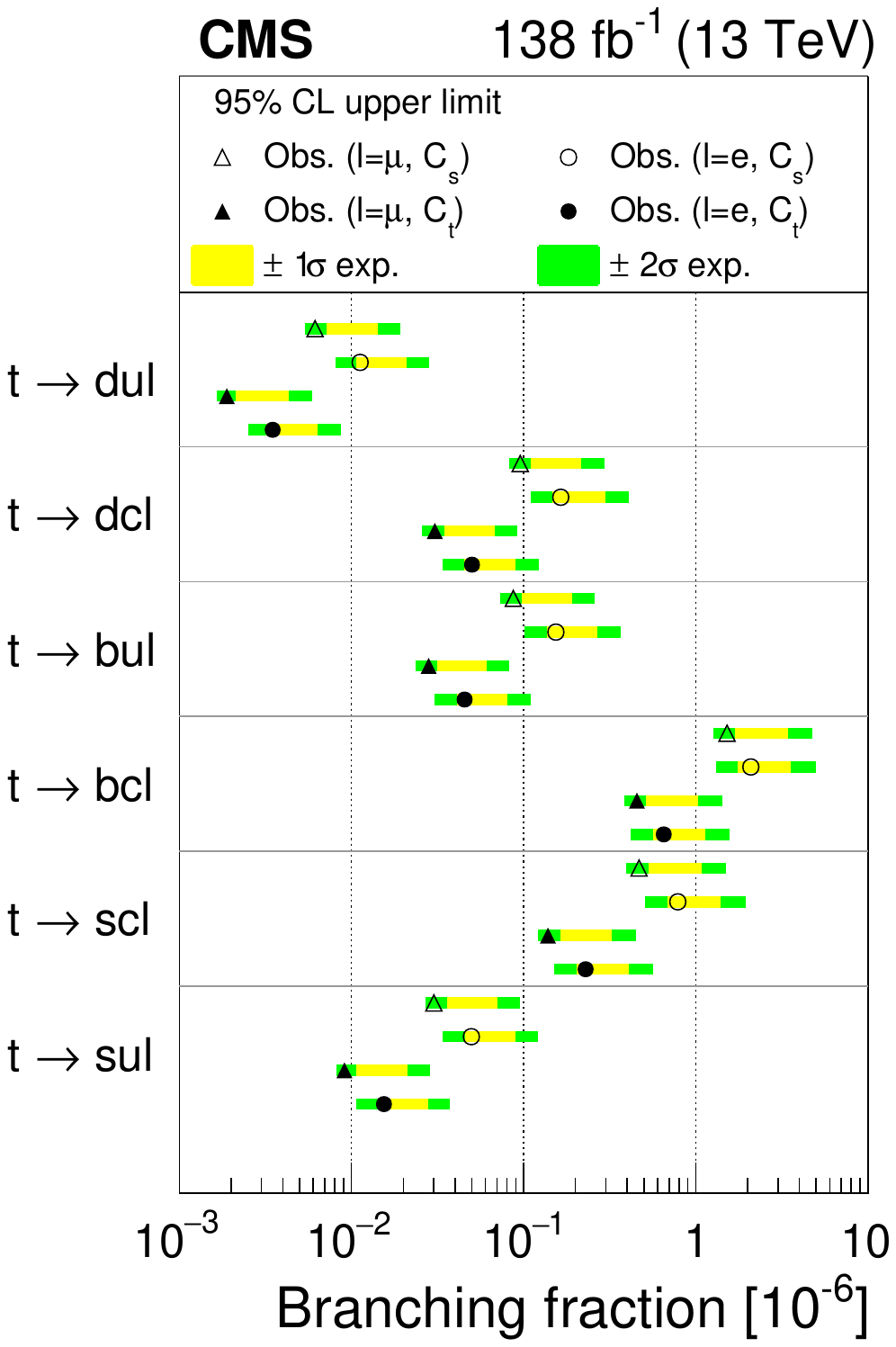}
  \caption{The observed upper limits on the branching fractions of the top quark BNV decays are shown with circle and triangle shapes for electron and muon couplings,  respectively.  
  The observed limits  corresponding to the $C_t$ and $C_s$ coefficients are shown with filled and open markers,  respectively.
  The  yellow light (green dark) bands indicate the range within plus or minus  one (two) standard deviations bands around the expected limits.
}
  \label{fig:summary}
  \end{center}
\end{figure}

Figure \ref{fig:BDT} shows the  BDT discriminant distributions  for events in the three channels (\ee, \emu,	 and \mumu) passing the event selection for the three data-taking years (2016--2018) combined. To illustrate signal distributions, simulated ``$\PQt \Pe \PQu\PQd$'' and ``$\PQt \PGm \PQu\PQd$'' samples are included in the figure, assuming $\Lambda=1\TeV$ and $C_t = C_s=1$. Signal events are well separated from background events.
To extract the signal contribution, a simultaneous  binned maximum-likelihood fit is performed of the BDT output distributions in the signal region for three years and three channels, with the systematic uncertainties described above treated as nuisance parameters. The best fit for the BNV effective couplings is consistent with zero and no significant excess over the background expectations is observed.
The sources of systematic uncertainty with the largest impact on the estimated signal
contribution depend on the  fermion flavor combination of the BNV interactions. The three main sources of uncertainty that are common among the BNV interactions are uncertainties in the normalization of the SM tW process, muon energy scale, and modeling of the top quark \pt  spectrum in the SM \ttbar simulation. 
The exclusion limits are calculated using the asymptotic approximation of the \CLs method \cite{Read:2002hq}. 
The adequacy of the asymptotic approximation has been validated with pseudo-experiments. 
The limit-setting procedure is performed for each individual BNV coupling  while setting the other BNV couplings to zero.
The observed and expected limits at 95\% confidence level (CL)  on the BNV effective coupling strengths are listed in Table \ref{Rfull}.
The limits on the strengths of the BNV couplings are translated to limits on the branching fractions for the BNV top quark decays.
The differences between different quark flavor combinations  stem mainly from the different PDFs involved in the production mode. 
The results for limits on various BNV  branching fractions are displayed in Fig. \ref{fig:summary}. 
Tabulated results are provided in the HEPData record for this analysis~\cite{hepdata}.

\begin{table}[htbp]
\centering
\topcaption{Expected and observed 95\% \CL upper limits on the BNV effective couplings and top quark BNV branching fractions. 
}
\label{Rfull}
\begin{scotch}{clllll}
 \rule{0pt}{2.6ex}
Vertex & $C_{\rm x}$  & $C_{\rm x}/\Lambda^2$   & $C_{\rm x}/\Lambda^2$ & $\mathcal{B}_{\rm x}$ & $\mathcal{B}_{\rm x}$ \\

     &           & [$\TeVns^{-2}$]      &[$\TeVns^{-2}$]  &   $[10^{-6}]$  & $[10^{-6}]$\\
         &         &Exp.              &Obs. &Exp.               &Obs.\\
\hline
\multirow{2}{*}{$\PQt\Pe\PQu\PQd$} & $s$ & 0.055 & 0.048 & 0.015 & 0.011   \\
 & $t$ & 0.031 & 0.027 & 0.005 & 0.003   \\
[\cmsTabSkip]
\multirow{2}{*}{$\PQt\PGm\PQu\PQd$} & $s$ & 0.046 & 0.036 & 0.010 & 0.006   \\
 & $t$  & 0.025 & 0.020 & 0.003 & 0.002   \\
[\cmsTabSkip]
\multirow{2}{*}{$\PQt\Pe\PQc\PQd$} & $s$ & 0.207 & 0.184 & 0.208 & 0.164   \\
 & $t$  & 0.114 & 0.102 & 0.063 & 0.050   \\
[\cmsTabSkip]
\multirow{2}{*}{$\PQt\PGm\PQc\PQd$} & $s$ & 0.178 & 0.141 & 0.153 & 0.095   \\
 & $t$  & 0.100 & 0.080 & 0.048 & 0.030   \\
[\cmsTabSkip]
\multirow{2}{*}{$\PQt\Pe\PQu\PQs$} & $s$ & 0.115 & 0.101 & 0.063 & 0.050   \\
 & $t$  & 0.064 & 0.056 & 0.019 & 0.015   \\
[\cmsTabSkip]
\multirow{2}{*}{$\PQt\PGm\PQu\PQs$} & $s$ & 0.102 & 0.079 & 0.050 & 0.030  \\
 & $t$  & 0.056 & 0.043 & 0.015 & 0.009   \\
[\cmsTabSkip]
\multirow{2}{*}{$\PQt\Pe\PQc\PQs$} & $s$ & 0.448 & 0.403 & 0.973 & 0.786   \\
 & $t$  & 0.243 & 0.218 & 0.286 & 0.229   \\
[\cmsTabSkip]
\multirow{2}{*}{$\PQt\PGm\PQc\PQs$} & $s$ & 0.394 & 0.311 & 0.752 & 0.468   \\
 & $t$  & 0.217 & 0.169 & 0.228 & 0.138   \\
[\cmsTabSkip]
\multirow{2}{*}{$\PQt\Pe\PQu\PQb$} & $s$ & 0.199 & 0.178 & 0.191 & 0.154   \\
 & $t$  & 0.109 & 0.097 & 0.057 & 0.045   \\
[\cmsTabSkip]
\multirow{2}{*}{$\PQt\PGm\PQu\PQb$} & $s$ & 0.168 & 0.134 & 0.136 & 0.087   \\
 & $t$  & 0.095 & 0.076 & 0.044 & 0.028   \\
[\cmsTabSkip]
\multirow{2}{*}{$\PQt\Pe\PQc\PQb$} & $s$ & 0.718 & 0.657 & 2.503 & 2.090   \\
 & $t$  & 0.405 & 0.367 & 0.795 & 0.652   \\
[\cmsTabSkip]
\multirow{2}{*}{$\PQt\PGm\PQc\PQb$} & $s$ & 0.703 & 0.564 & 2.393 & 1.521   \\
 & $t$  & 0.386 & 0.307 & 0.722 & 0.455   \\
\end{scotch}
\end{table}

In summary,  a search for baryon number violation (BNV) in events with top quarks is performed using the  LHC proton-proton collision data at a center-of-mass energy of 13\TeV.  The analysis explores baryon number violating effects in single top quark production for the first time. Data were collected by the CMS experiment in 2016--2018 and  correspond to an integrated luminosity of 138 \fbinv. Events with a lepton pair and exactly one \PQb-tagged jet are selected. A boosted decision tree (BDT) is used to separate signal events from background events. A binned maximum likelihood fit to the BDT output distribution is performed to search for the BNV processes.  Considering BNV vertices in the production of top quarks dramatically increases the sensitivity of this search.
No significant excess of events over the background prediction is observed.  Upper limits are placed on the strengths of the BNV couplings, which are multiple orders of magnitude more stringent than the previous limits \cite{CMS:2013zol}.  

\begin{acknowledgments}
We congratulate our colleagues in the CERN accelerator departments for the excellent performance of the LHC and thank the technical and administrative staffs at CERN and at other CMS institutes for their contributions to the success of the CMS effort. In addition, we gratefully acknowledge the computing centers and personnel of the Worldwide LHC Computing Grid and other centers for delivering so effectively the computing infrastructure essential to our analyses. Finally, we acknowledge the enduring support for the construction and operation of the LHC, the CMS detector, and the supporting computing infrastructure provided by the following funding agencies: SC (Armenia), BMBWF and FWF (Austria); FNRS and FWO (Belgium); CNPq, CAPES, FAPERJ, FAPERGS, and FAPESP (Brazil); MES and BNSF (Bulgaria); CERN; CAS, MoST, and NSFC (China); MINCIENCIAS (Colombia); MSES and CSF (Croatia); RIF (Cyprus); SENESCYT (Ecuador); ERC PRG, RVTT3 and MoER TK202 (Estonia); Academy of Finland, MEC, and HIP (Finland); CEA and CNRS/IN2P3 (France); SRNSF (Georgia); BMBF, DFG, and HGF (Germany); GSRI (Greece); NKFIH (Hungary); DAE and DST (India); IPM (Iran); SFI (Ireland); INFN (Italy); MSIP and NRF (Republic of Korea); MES (Latvia); LMTLT (Lithuania); MOE and UM (Malaysia); BUAP, CINVESTAV, CONACYT, LNS, SEP, and UASLP-FAI (Mexico); MOS (Montenegro); MBIE (New Zealand); PAEC (Pakistan); MES and NSC (Poland); FCT (Portugal); MESTD (Serbia); MCIN/AEI and PCTI (Spain); MOSTR (Sri Lanka); Swiss Funding Agencies (Switzerland); MST (Taipei); MHESI and NSTDA (Thailand); TUBITAK and TENMAK (Turkey); NASU (Ukraine); STFC (United Kingdom); DOE and NSF (USA).
\end{acknowledgments}

\bibliography{auto_generated}

\cleardoublepage \appendix\section{The CMS Collaboration \label{app:collab}}\begin{sloppypar}\hyphenpenalty=5000\widowpenalty=500\clubpenalty=5000
\cmsinstitute{Yerevan Physics Institute, Yerevan, Armenia}
{\tolerance=6000
A.~Hayrapetyan, A.~Tumasyan\cmsAuthorMark{1}\cmsorcid{0009-0000-0684-6742}
\par}
\cmsinstitute{Institut f\"{u}r Hochenergiephysik, Vienna, Austria}
{\tolerance=6000
W.~Adam\cmsorcid{0000-0001-9099-4341}, J.W.~Andrejkovic, T.~Bergauer\cmsorcid{0000-0002-5786-0293}, S.~Chatterjee\cmsorcid{0000-0003-2660-0349}, K.~Damanakis\cmsorcid{0000-0001-5389-2872}, M.~Dragicevic\cmsorcid{0000-0003-1967-6783}, P.S.~Hussain\cmsorcid{0000-0002-4825-5278}, M.~Jeitler\cmsAuthorMark{2}\cmsorcid{0000-0002-5141-9560}, N.~Krammer\cmsorcid{0000-0002-0548-0985}, A.~Li\cmsorcid{0000-0002-4547-116X}, D.~Liko\cmsorcid{0000-0002-3380-473X}, I.~Mikulec\cmsorcid{0000-0003-0385-2746}, J.~Schieck\cmsAuthorMark{2}\cmsorcid{0000-0002-1058-8093}, R.~Sch\"{o}fbeck\cmsorcid{0000-0002-2332-8784}, D.~Schwarz\cmsorcid{0000-0002-3821-7331}, M.~Sonawane\cmsorcid{0000-0003-0510-7010}, S.~Templ\cmsorcid{0000-0003-3137-5692}, W.~Waltenberger\cmsorcid{0000-0002-6215-7228}, C.-E.~Wulz\cmsAuthorMark{2}\cmsorcid{0000-0001-9226-5812}
% ...
\par}
\vskip\cmsinstskip
\dag:~Deceased\\
$^{1}$Also at Yerevan State University, Yerevan, Armenia\\
$^{2}$Also at TU Wien, Vienna, Austria\\
% ...
\end{sloppypar}
%%% END EDITABLE REGION %%%
% skeleton_end
\end{document}